\documentclass{article}
\usepackage[a4paper]{geometry}
\usepackage{authblk}
\usepackage{hyperref}
\usepackage{amsfonts}
\usepackage{url}
\usepackage{booktabs}
\usepackage{dcolumn}

\usepackage{cite}
\usepackage{hyperref}
\hypersetup{
    colorlinks=true,
    linkcolor=black,
    filecolor=magenta,      
    urlcolor=cyan,
    citecolor=blue,
}
\usepackage{amsmath}
\usepackage{mathtools}
\usepackage{mathdots}

\usepackage{float}

\begin{document} 
\title{GAPS: Geo Data Portals for Air Pollution Studies}

\author[1]{Filipe Fernandes}
\author[2]{Helena Serrano}
\author[2]{Maria Alexandra Oliveira}
\author[2]{Cristina Branquinho}
\author[1]{João Nuno Silva}

\affil[1]{INESC-ID, Instituto Superior Técnico, Universidade de Lisboa, 1000-029 Lisboa, Portugal}
\affil[2]{Centre for Ecology, Evolution and Environmental Changes, Faculdade de Ciências da Universidade de Lisboa, 1749-016 Lisboa, Portugal}
\affil[ ]{\textit {filipe.m.m.fernandes@tecnico.ulisboa.pt, \{hcserrano, maoliveira, cmbranquinho\}@fc.ul.pt,  joao.n.silva@inesc-id.pt}}

\maketitle

\thispagestyle{plain}
\pagestyle{plain}

\begin{abstract}
There is a wealth of data on air pollution within several users' reach, including modelled concentrations and depositions as well as observations from air quality stations. However, data integration to perceive spatial and temporal trends at the national level is a complex undertaking. The difficulties are mainly related to the data sources (many files with a lot of information and in different formats). In addition, the processing of this data is time-consuming and impractical when the time period under analysis increases.
Furthermore, the processing of spatial-temporal data requires the use of multiple specific tools, such as geographic information systems software
and statistical,
which are not readily accessible to non-specialised users. 
The proposed solution is to develop libraries that are responsible for aggregating different types of data related with pollution. In addition to allowing the integration of different data, the libraries are also the basis for developing web applications. The libraries allow the collection of data made available by the Portuguese Environment Agency (APA) and official modelling results for Europe. The applications developed will extend the libraries to allow data processing and representation through Dashboards. These Dashboards provide access to non-specialised users, namely regarding the trends of pollutants in each region.
The implemented libraries allow the development of projects that would need access to the same data sources. These Dashboards allow for simplified access to data and for studies that have so far been beyond the reach of researchers.
\end{abstract}


\section{Introduction}
%
%
%
%
\label{chapter:introduction}

Air pollution is a global problem recognised internationally (e.g., UN Sustainable Development Goals of the United Nations \cite{sustainable-development-goals}) that affects both the environment and human health. Because of these harmful effects, there is a considerable investment made by the European Commission and member states in determining the concentration/deposition of pollutants, the cause-and-effect relationship between the concentration of pollutants and the problems it can cause to the environment and to living beings \cite{directive2016directive}.

To better understand these cause-and-effect relationships it is necessary to have access to spatially and temporally explicit information, preferably in the form of observations (certified measurements from air quality stations) of pollutants. However, due to the limited number and unrepresentative character of air quality stations \cite{YATKIN2020225}, alternative information must be used, such as pollution concentration/deposition provided by air pollution chemical transport models \cite{OLIVEIRA2020117290}. These models take into account meteorological conditions (e.g., wind direction and speed, precipitation) as well as the main known emission sources at different heights, and simulate chemical reactions in the atmosphere, pollutant concentrations and deposition in different landcovers \cite{acp-12-7825-2012}. As a result, models provide the spatial distribution of concentration and deposition for different pollutants, and their change in time.
In Europe, major concerns regarding air pollution are addressed in the EU Directive 2016/2284 \cite{directive2016directive}, and mostly focus on nitrogen, sulphur, particles and ozone. European air policy assessments rely on data provided by the EMEP/MSC-W (European Monitoring and Evaluation Programme/Meteorological Synthesizing Center-West) transport model \cite{acp-12-7825-2012}, herein referred to as EMEP model.

For a model to be used by decision makers, it must first be evaluated. This evaluation is usually based in a comparison between observations and model predictions by using a set of statistical measures that determine bias, data scatter and error \cite{chang2004air}.

In Portugal, air pollution concentrations are measured in air quality stations and made available by the environmental Portuguese agency, \textit{Agência Portuguesa do Ambiente} (APA). Although, these stations do not fully represent air pollution throughout Portugal, the observations can be used to evaluate model results. The APA data is stored in tabular form, in the old version excel format, .xls. Regardless, each excel file can represent one station, containing the respective pollutants or it can represent one pollutant, containing the values of the stations where it was captured. So there are many different ways to retrieve and store this data. The size of each excel varies, but normally is less than 1MB, nonetheless the information is divided in multiple excels. 

Model results used in this work are provided by EMEP and extracted for mainland Portugal. EMEP model results are provided as a raster (array of geographical pixel values) in NetCDF format. The EMEP files can have very different sizes, ranging from 70 MB to 20 GB, depending on the temporal resolution (year, month, day, and hour) and the number of pollutants. 


To complement, the observation and simulation data, it is useful to provide information about the spatial distribution of land-use, or ecosystem . An ecosystem is a community of living organisms in conjunction with the nonliving components of their environment. Air pollution affects ecosystems and by knowing what type of ecosystem exists at a particular location, it is possible to determine whether or not pollutants cause harmful effects at that location.

For land use, COS data set \cite{COS_especificacoes, CosDownload}, provides high resolution information in GIS vector file format (geometrical shapes often associated with a tabular database that describes their attributes) within the Portuguese territory. MAES data set \cite{MAES_website, MAESDownload} provides high resolution information about the type of ecosystem in a raster file format (each pixel has an index value that, with a table, identifies the type of ecosystem) within the European territory and it has a maximum resolution of 100 m x 100 m.

There are challenges to aggregate all this data. First of all, there are many data sources that work and store data differently. In addition, the files are of complex types, and can be large in size or large in quantity. Therefore, data processing from aggregation, to spatial overlap and comparison, also differs for each source.

The objective of this work is to develop libraries and applications to retrieve, process, aggregate and present spatial and temporal data of air pollution. With the integration of the developed libraries, a set of demonstration applications was developed that solves some of the problems of data handling, overlap and visualisation of non-specialised users.

The paper is structured as follows: 
Section II introduces the infrastructure/tools already available.
Section III describes the requirements for a programmer to work with pollution data, architecture, and the libraries that have been developed. 
Section IV describes the applications developed using the libraries presented in Section III. 
Finally, Section V summarises the main results and conclusions.

\section{Related Work}
\label{sec:backg}

\subsection{Data Formats}
\label{subsection:data_formats}


The data can be stored in many different formats.

The vector format allows to represent geographic features such as points, lines and polygons with great precision. Generally, each point is represented as a single coordinate pair, while lines and polygons are represented as ordered lists of vertices/points. Attributes are associated with each vector feature \cite{GIS_dictionary}. This spatial data format is used preferably in applications with geographic limits.

The raster format allows to store spatial information in a grid format, where the locations are represented by an array of cells (pixels) \cite{raster_data}. It is used, for example, to represent model results.

An ESRI shapefile is a vector data storage format, for storing shapes and attributes of geographic features \cite{GIS_dictionary}. It is stored in a set of related files and contains one feature class.

The Network Common Data Form (NetCDF) is a file format for storing multidimensional data \cite{netcdf_data}. This format is divided in two parts \cite{netcdf_data_unidata}: a header that contains all the information about dimensions, attributes, and variables; a data part that comprises, fixed-size data, containing the data for variables that don't have an unlimited dimension, and variable-size data, containing the data for variables that have an unlimited dimension.

Tagged Image File Format (TIFF) is an industry standard format for handling raster or bitmapped images, that can be saved in a variety of colours \cite{Tiff-definition}. Based on TIFF format, there is a similar format used for georeferenced raster imagery, named GeoTIFF \cite{GeoTiff-definition}.

\subsection{Geographical Information System}
\label{subsection:GIS}
Geographical Information System (GIS) is a framework to aggregate, manage and analyse different types of data, namely spatial data. By using this tool, it is possible to analyse geographic data and organise it in different layers of information to later visualise as maps \cite{esri-what-is-gis}. With GIS, the spatial data are no longer represented in image/painting format, as in the conventional maps (paper map), but are now represented as digital information, providing a higher flexibility in data representation. The data can be represented as an image containing cartographic information (e.g., roads, urban/forest/agricultural areas, ecosystems), or as a set of statistical tables, that later can be converted into graphical content, such as scatter plots. The data is saved in a structured format which makes it easier to collect, analyse and save \cite{whatgis, what-is-gis}.

As the data, whether spatial or not, is stored in a database, it can be related. That is, it is feasible to relate spatial data with non-spatial data (e.g., tabular data) through queries to generate maps or to obtain textual attributes.

The data with geographic attributes can be represented by two different formats, vector and raster. 

\subsection{Data Storage}
\label{subsection:Database}
The data that can be saved on the server, can have several formats, which can be stored in tables or in files using two separate models. 

Spatial DataBase Management Systems (spatial DBMS) although not developed as a software solution, it offers the ability to store cartographic data in DBMS. One of the possible DMBS to use is PostgreSQL, but this service lacks the capability to work with data that have geographical attributes. PostGIS is a plugin that allows to save spatial data types, as well as, functions for analysis and for processing. This plugin is used to expand the PostgreSQL database by providing the ability to save spatial data on a open-source object-relational database management system \cite{steiniger2012free, postgresql, postgis}.


GeoServer is a three-tier client-server architecture consisting of a Web Server, a Web GIS software, and a database. GeoServer allows to share and edit geospatial data. The system was designed to have interoperability, allowing to publish data of any spatial type, making any geospatial information available as much as possible,  \cite{mehdi2014implement}.

\begin{figure}[!htb]
    \centering
    \includegraphics[width=7.5cm]{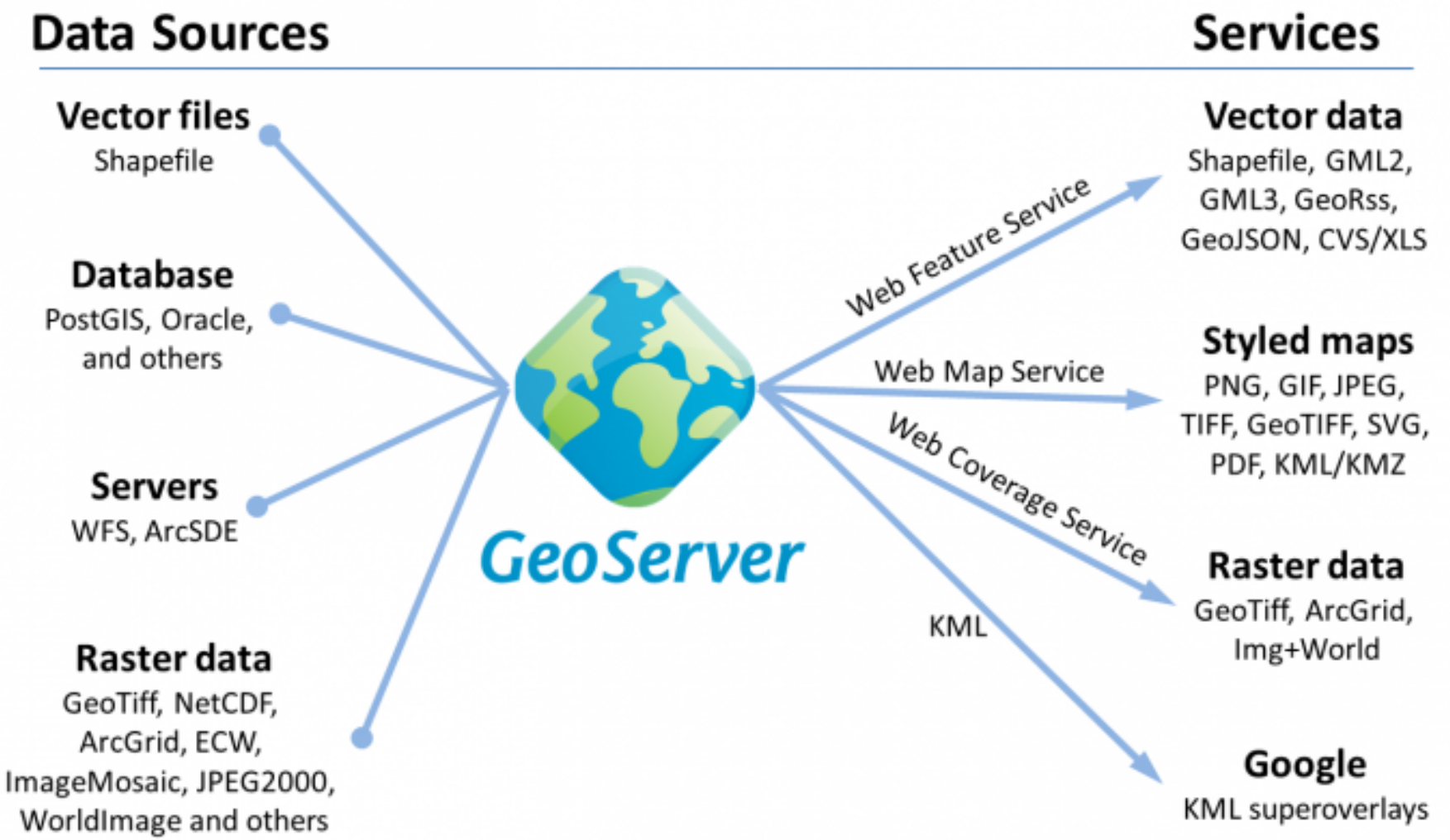}
    \caption[GeoServer Diagram]{GeoServer Diagram \cite{eatlas}.}
    \label{fig:geoserver_diag}
\end{figure}

GeoServer, besides being able to save data, is also used to publish data, meaning it allows exposing spatial information to any user. This exposure is possible through the implementation of Open Geospatial Consortium (OGC) standards. The OGC standards consist of more than 30 standards that are responsible to cover services, which include standards for distributing spatial and tabular data. There are also standards that are responsible to edit data style. Web Map Service (WMS), Web Feature Service (WFS), and Web Coverage Service (WCS) are important OGC standards \cite{what-is-Geoserver}. With WMS it is possible to access and create maps in different output formats. On the other hand, through the WFS and WCS standards it is possible to share and edit the style of the data that is used to generate the maps, and for the WFS standard it is also possible to edit the data content. WCS is used for data in raster format, while WFS works with vector formats. Through its standards, GeoServer is thus a service that bridges the gap between the various sources of spatial data and the different services for each type of data (figure \ref{fig:geoserver_diag}).

Therefore, through its architecture, GeoServer is used as a database and to create maps from the data that is stored on the system. With the OGC standards, it is easy to share the data with any kind of system, and also it is easy to perform queries to obtain the desired information from the stored data.

\subsection{Data Management}
\label{subsection:GeoNode}

Spatial Data Infrastructures (SDI) is a framework that analyses cartographic data and has an application for users and suppliers to communicate. It also facilitates the access to the geographic data because it uses a minimal set of practices, protocols and standards \cite{SDI_definition}.

GeoNode is a web-based platform that implements GIS and SDI through an open-source framework. As interoperability is one of the main concepts of SDI, it implies that GeoNode is based on technologies with standards provided by OGC. Thus, GeoNode offers the following features: spatial search for data and metadata;  management and sharing of data as well as the management of sharing policies; data visualisation as well as the integration of different sources that are stored in external infrastructures and services. The visualisation is accessed through WMS standards. 

This technology is a web application developed in Django. Django is a high-level Python Web framework, which allows to easily integrate new modules (applications), and also to access resources from the web.

GeoNode is able to provide a user interface using the Django templates and JavaScript libraries / frameworks. One of the frameworks used is jQuery which is used, for example, in search boxes, and to provide dynamism to the web page. Ajax is another technology that is used to make requests to the application during the interaction with the client. For example, Ajax is used to obtain the most recent data stored in the server and to print this result on the page. Finally, this user interface includes technology based on OpenLayers or Leaflet to create interactive / dynamic maps.

Because GeoNode is a web application, it is possible to run on a web server, such as, nginx or Apache. With a web server it is possible to make the content available to anyone that can access the Internet.


\section{Libraries for pollution data programming}
\label{sec:imple}

\subsection{Requirements}
\label{section:library:requirements}

The requirements indicate what will be possible to do with the libraries. These requirements are necessary for the programmer to work with data pollution. If these requirements are all implemented then there will be a set of libraries that can help to create web applications that facilitate the researchers in their studies. 

The proposed libraries' requirements are:

\begin{enumerate}
    \item The library allows the development of applications that use georeferenced data stored in excel;
    
    \item The library allows the collection and processing of pollution data from the Portuguese network of air pollution stations;
    
    \item The library allows EMEP model predictions data management;
    
    \item The library allows an easy access to Geonames service;
    
    \item Libraries allow the management of equivalences between EMEP model predictions and observations data;
    
    \item Libraries must run as part of a Middleware for web/georeferenced application development;
    
    \item Libraries must store data in geospatial database;
    
    \item Libraries must store data on a geospatial data sharing server;
    
    \item Libraries must use a standard API for database access;
\end{enumerate}

\subsection{Architecture}
\label{section:library:Architecture}

The system must be able to integrate the requirements announced in section \ref{section:library:requirements}, namely allow easy integration of libraries. In addition, it must be as generic as possible in order for any programmer to be able to work with the system.

The framework provides a generic functionality that can be changed by additional user-written code (programmer), thus providing application-specific software. In addition, it has a standard and universal way to build and deploy applications, and it may include support code libraries that bring together all the different components. According to the requirements, the system must be able to support communication with other systems via Ethernet, and must be web-based.

The required data storage must be able to store data in tabular and file formats. The Database is an additional system that allows to store data in table format, and the Spatial Data Server is responsible to manage spatial data, including data that is not possible to store in a table.

\begin{figure}[!ht]
    \centering
    \includegraphics[width=7.5cm]{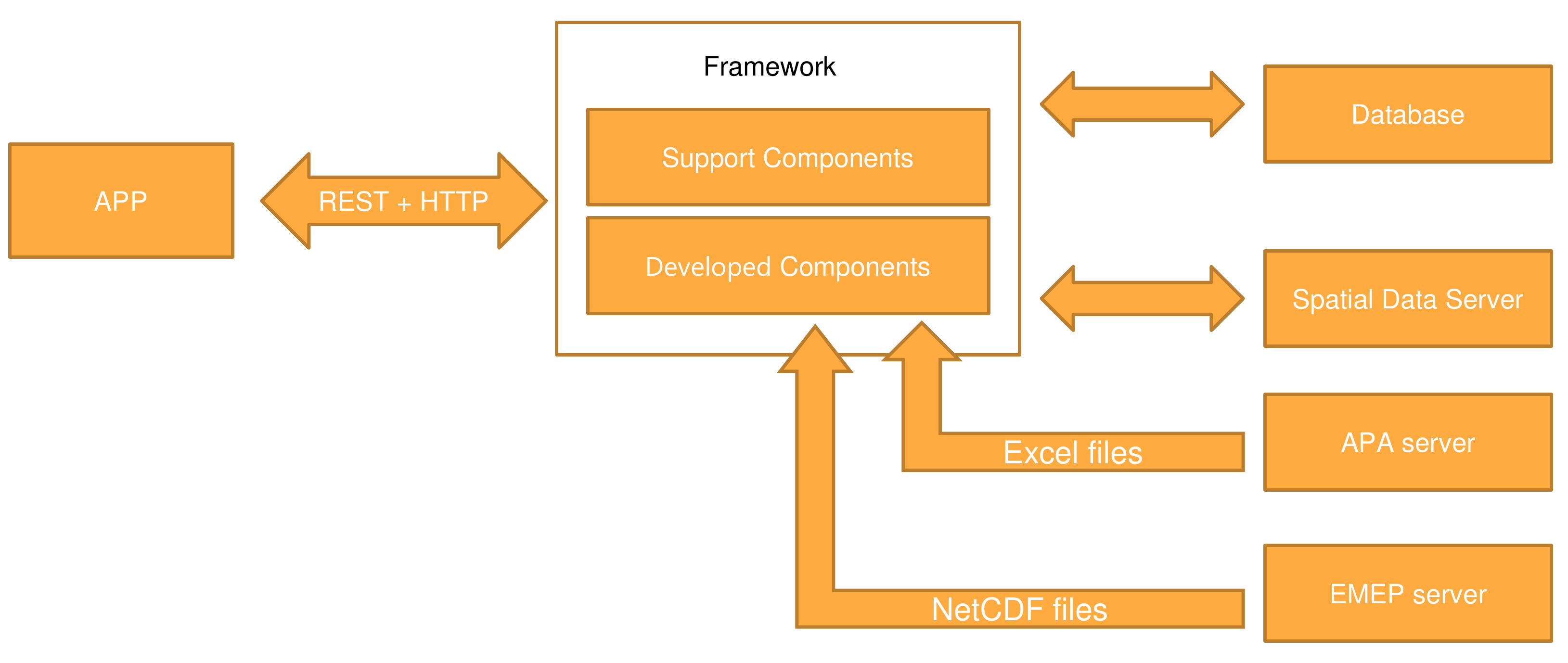}
    \caption{Generic Architecture.}
    \label{fig:General_arch}
\end{figure}

Figure \ref{fig:General_arch} shows a generic representation of the architecture of a system, that is capable to integrate support and user-written code. Due to the framework, it is possible to connect to different storage systems and, at the same time, to develop applications that facilitate the access to results, which were previously complicated to access. The system is also capable to communicate with observations and model predictions servers.

The generic architecture, presented in figure \ref{fig:General_arch}, is already implemented in GeoNode, a product that it is available to the public. The GeoNode is implemented in Python and provides a web-based platform that implements GIS and SDI.

Since GeoNode is already linked to a Database and a Spatial Data Server, there is no need to develop a new platform, because it is possible to change the behaviour of the GeoNode system by implementing modules/libraries.

Even though, it is possible to work with any systems for the Database and the Spatial Data Server, these are based on the GeoNode product. The chosen Database, PostgreSQL, is a database management system that, together with the PostGIS plugin, is capable of storing data with geographical attributes. GeoServer is the chosen Spatial Data Server.

\begin{figure}[!ht]
    \centering
    \includegraphics[width=7.5cm]{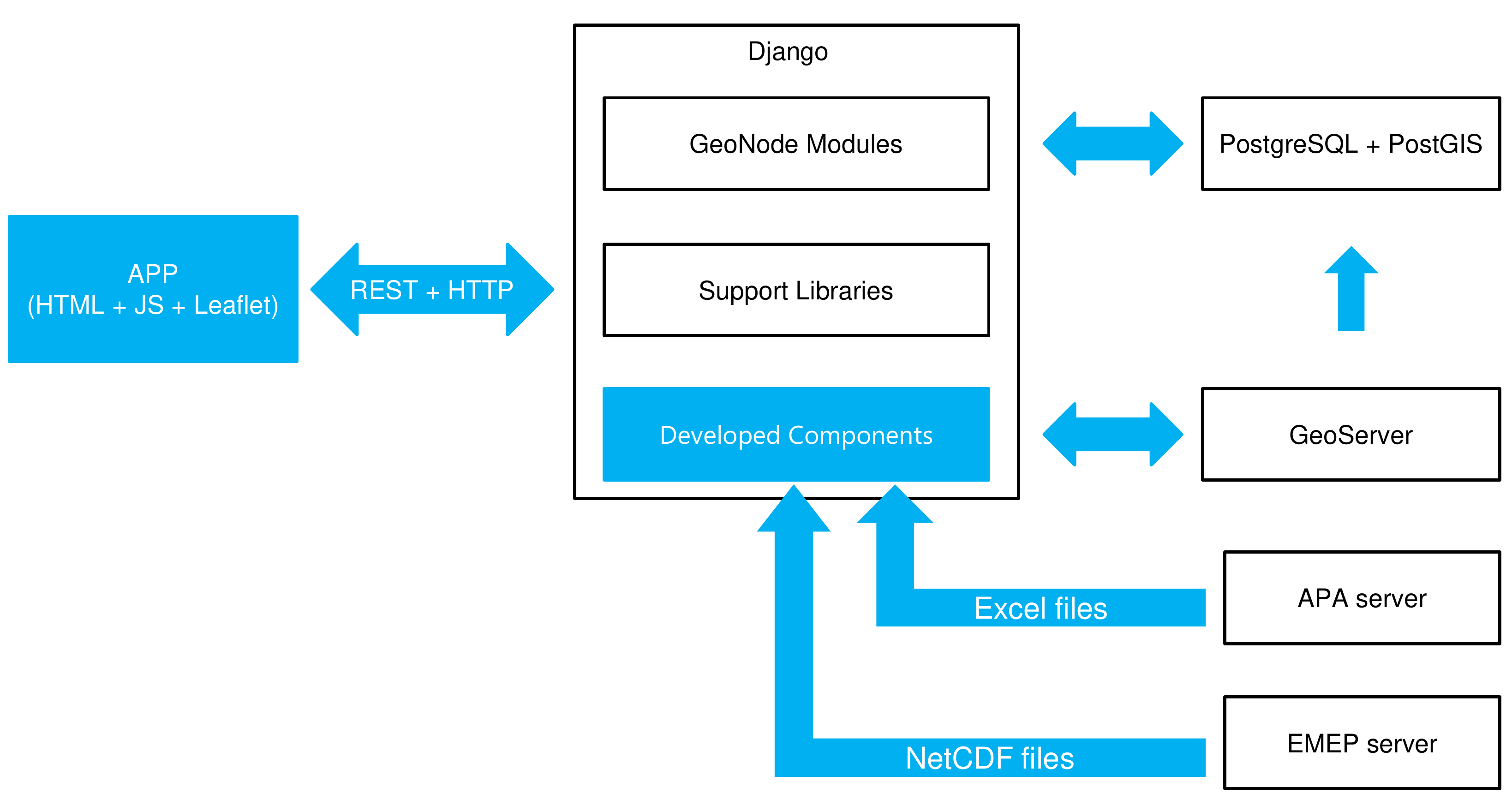}
    \caption{Architecture with the supporting systems. Blue boxes can be developed by the programmer, and the black bounded boxes have been developed by other entities.}
    \label{fig:Instancead_arch}
\end{figure}

Figure \ref{fig:Instancead_arch} shows a detailed version of the system architecture where the systems/libraries used are instantiated. The developed components are responsible for managing  observations and model predictions and, therefore, communicate with data source servers. The developed system takes advantage of some modules of the GeoNode, namely the graphical interface for the submission of georeferenced data. Therefore, this architecture is based on the GeoNode product but only uses some of its modules.

GeoNode modules do not cover all the programming needs to implement the requirements. For this reason other support libraries must be added, such as: NetCDF4 \cite{netcdf4}; Pandas \cite{pandas}; numpy \cite{numpy}; rasterio \cite{rasterio}; shapely \cite{shapely}; fiona \cite{fiona} and Leaflet \cite{leafletjs}.

\subsection{GeoExcel}
\label{section:library:Geo-Excel}

The objective of this component is to facilitate the integration of georeferenced data in excel format to use in future applications. GeoExcel is an extension of the GeoNode component due to its limitations.

\begin{figure}[!ht]
    \centering
    \includegraphics[width=7.5cm]{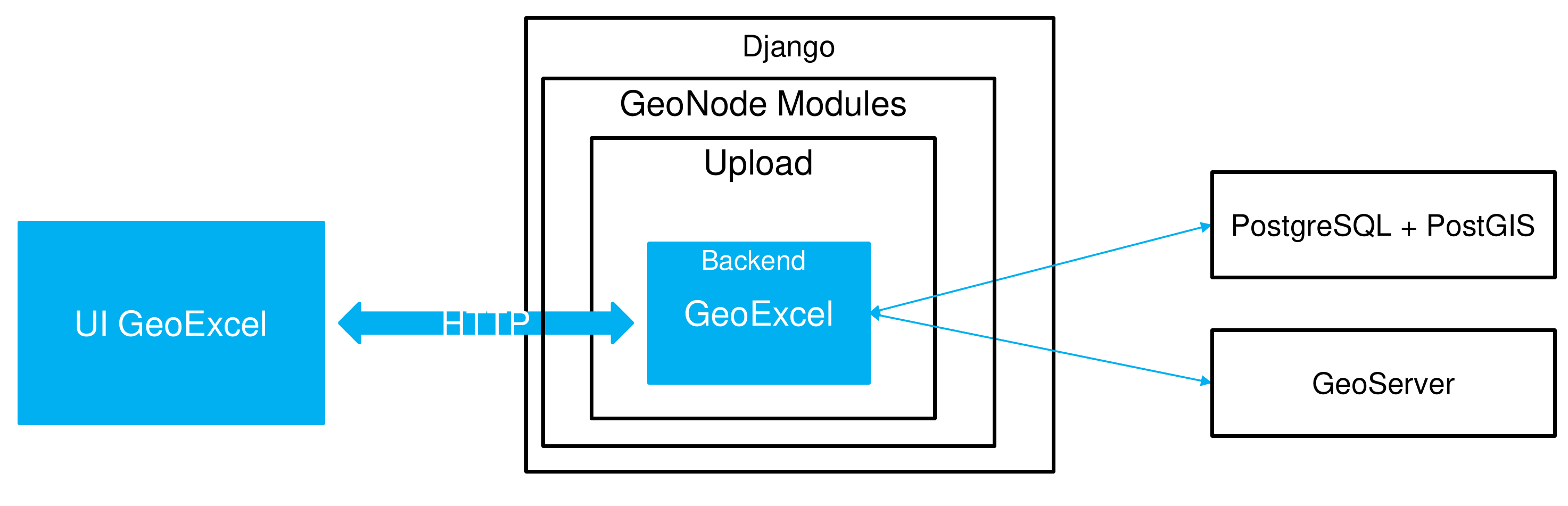}
    \caption{GeoExcel component in a generic Django framework. Blue boxes corresponds to GeoExcel, and the black bounded boxes have been developed by other entities.}
    \label{fig:GeoExcel_diagram}
\end{figure}

GeoNode has a graphical interface where the user can upload files with spatial attributes to the GeoServer, like shapefiles and geotiffs. The GeoNode module responsible for uploading data to the GeoServer is named Upload. Figure \ref{fig:GeoExcel_diagram} represents a diagram of the integration of GeoExcel in GeoNode modules. The GeoExcel is inside of Upload Module because it is an extension that allows the upload of excel files. Therefore, GeoExcel was developed as an extension of GeoNode that allows the upload of excel files.

When the GeoNode module detects that the uploaded file is an excel file, it runs the GeoExcel code. This code does the following procedures. Firstly, the excel file must be opened to access the data. Pandas is a library used to control and to access excel files. After access is established, the next step is to verify if the table exists in the database, otherwise it must be created. The final step is to copy the data from the excel file to the table in the database. As the final step depends on the type of data that is stored in the excel file, the process calls a function responsible for parsing the excel file, that is an input parameter of the method.

\begin{figure}[!ht]
    \centering
    \includegraphics[width=7.5cm]{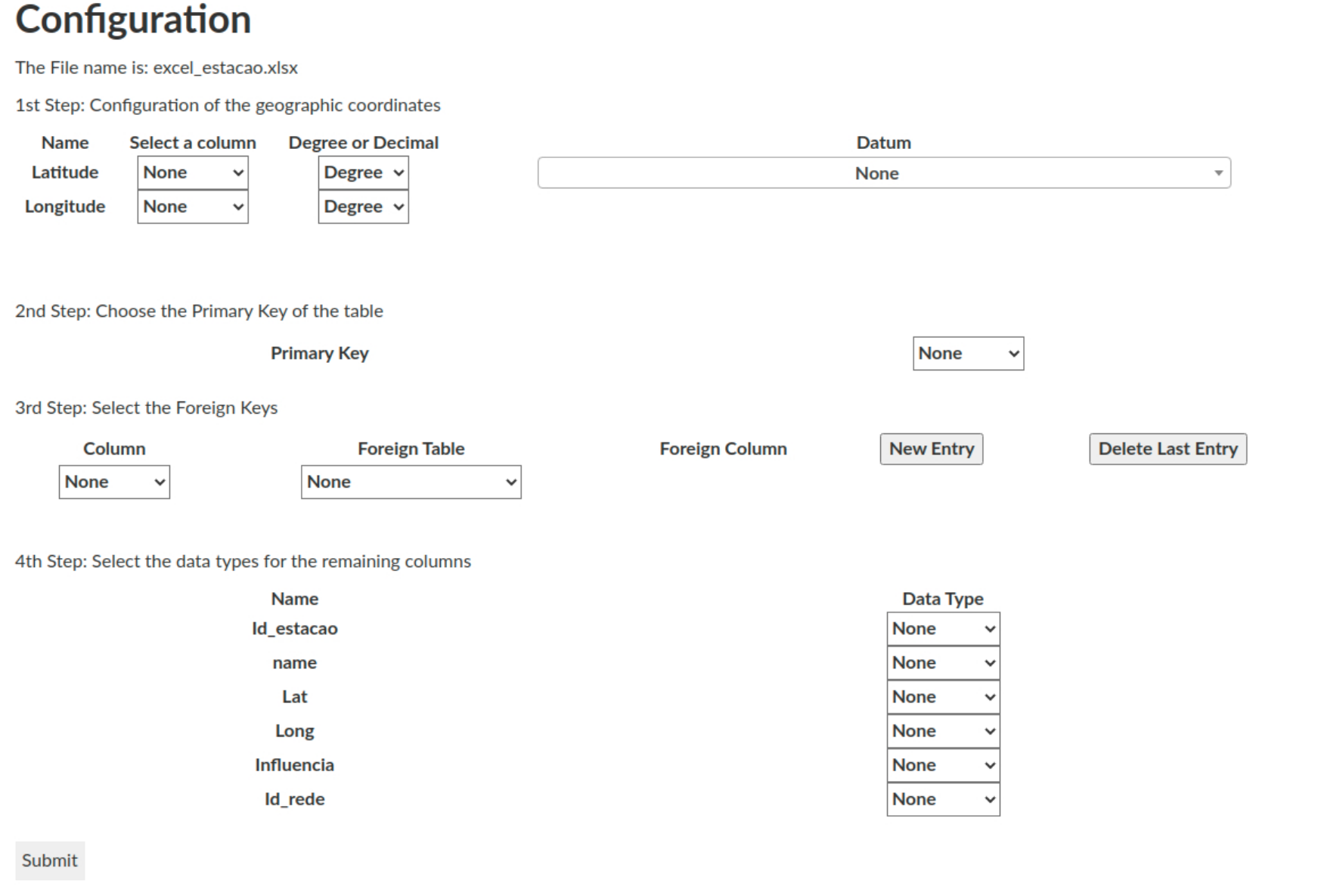}
    \caption{GeoExcel user interface, where the user can configure the parser to upload the excel to the database.}
    \label{fig:Configurador_excel}
\end{figure}

User Interface (UI) is an extension of the Upload module UI, that allows to the user to upload excel files. After choosing an excel file, the user will have to choose either to create a new table or to update an existing one. If the user wants to update a table, then has to choose the table, being redirected to the initial web page that is the starting point to upload any file. If the user wants to upload the information to a table, then has to choose the name of the table and after that the user is redirected to a web page where the user must configure the parser that is used to upload the excel file (figure \ref{fig:Configurador_excel}). 

This configuration (figure \ref{fig:Configurador_excel}) is divided in four steps: if the excel contains coordinates, then the user must select which columns represent latitude and longitude, the format of the coordinates, and the coordinate system; the second step is to choose the table's primary key; the third step is to connect the new table to another tables using foreign keys; the last step is to choose the data type (Integer, Float, String and Date/Time) of the remaining columns (the column that represents the coordinates and the columns that are linked to foreign tables already have the type of data defined).

The GeoExcel component developed during this project is available in a git repository, \href{https://github.com/FMMFHD/GeoExcel}{https://github.com/FMMFHD/GeoExcel}.

\subsection{WebAPA}
\label{section:library:WebAPA}

The objectives of the WebAPA component are to store observations data from the APA server in a georeferenced database. Besides storage, it offers an API for developing web applications with geospatial data.

\begin{figure}[!ht]
    \centering
    \includegraphics[width=7.5cm]{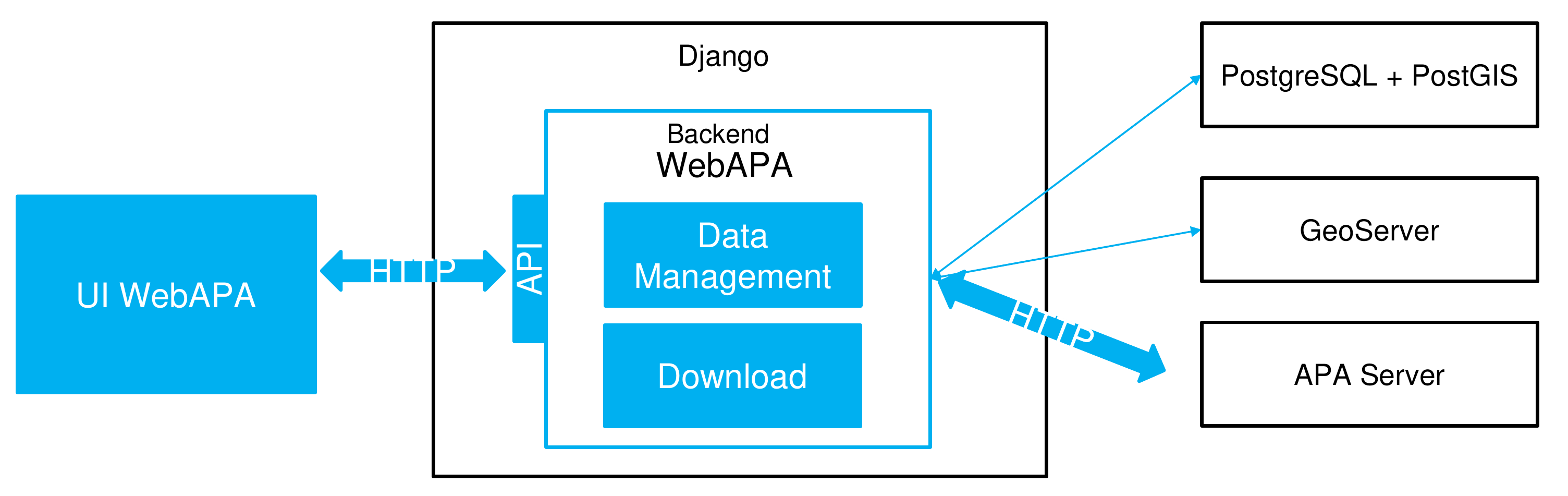}
    \caption{WebAPA component in a generic Django framework. Blue boxes corresponds to WebAPA, and the black bounded boxes have been developed by other entities.}
    \label{fig:WebAPA_diagram}
\end{figure}

WebAPA architecture is illustrated in figure \ref{fig:WebAPA_diagram}. The backend part is responsible to get and manage the data. The User Interface allows the user to download the data for all available years.

There are some problems with the data downloaded from the APA server. The data is accessed by HTTP, because there are no web services to collect this data. To download the excel file it is necessary to select several inputs, because the downloaded data from APA is stored in excels files, one file per station/year. The only way to download the data is to replicate the interactions of the user through HTTP, using web scrapping, a technique used for extracting data from websites. To store the downloaded files, a method from the GeoExcel module is used to upload the excel files to the database. These files do not have information about the station location. So, it is necessary to get the locations of each station and then relate that information with pollutant concentrations from the air quality stations.

The data management is simple because the data is stored in the database. This data is only replaced by the system administrator that uses the user interface. To access the data, it is only necessary to create SQL statements to filter data that is stored in the database.

The data is divided by years at the APA server. Therefore, the objective of the user interface is to allow the admin user to select a year to download the data from the APA server. This process is time-consuming because it is necessary to download many files. The communication channel between the interface and the server is not indefinitely open. Thus, this interface has a limitation, because the user might never receive a success or error message, not knowing whether the execution was a success or not. To confirm that the data has been downloaded, the interface provides an input text and a button that allows this verification.

API is an interface that the programmer can use to obtain data. The available access points allow the programmer to obtain: a list of all the pollutants measured in a given station; a list of all the observations for a given station, pollutant, date, and temporal resolution.

The WebAPA component developed during this project is available in a git repository, \href{https://github.com/FMMFHD/WebAPA}{https://github.com/FMMFHD/WebAPA}.

\subsection{WebEMEP}
\label{section:library:WebEMEP}

The objectives of the WebEMEP component are to store all the available model predictions from the EMEP server in a local database. Besides storage, it offers access mechanisms and data management for a framework where web applications with geospatial data are developed. However, there are some issues concerning the data that contains the EMEP model results. The data stored in the EMEP server is accessed by HTTP through a catalogue. To collect the data it is necessary to consult the catalogue and parser it. 

\begin{figure}[!ht]
    \centering
    \includegraphics[width=7.5cm]{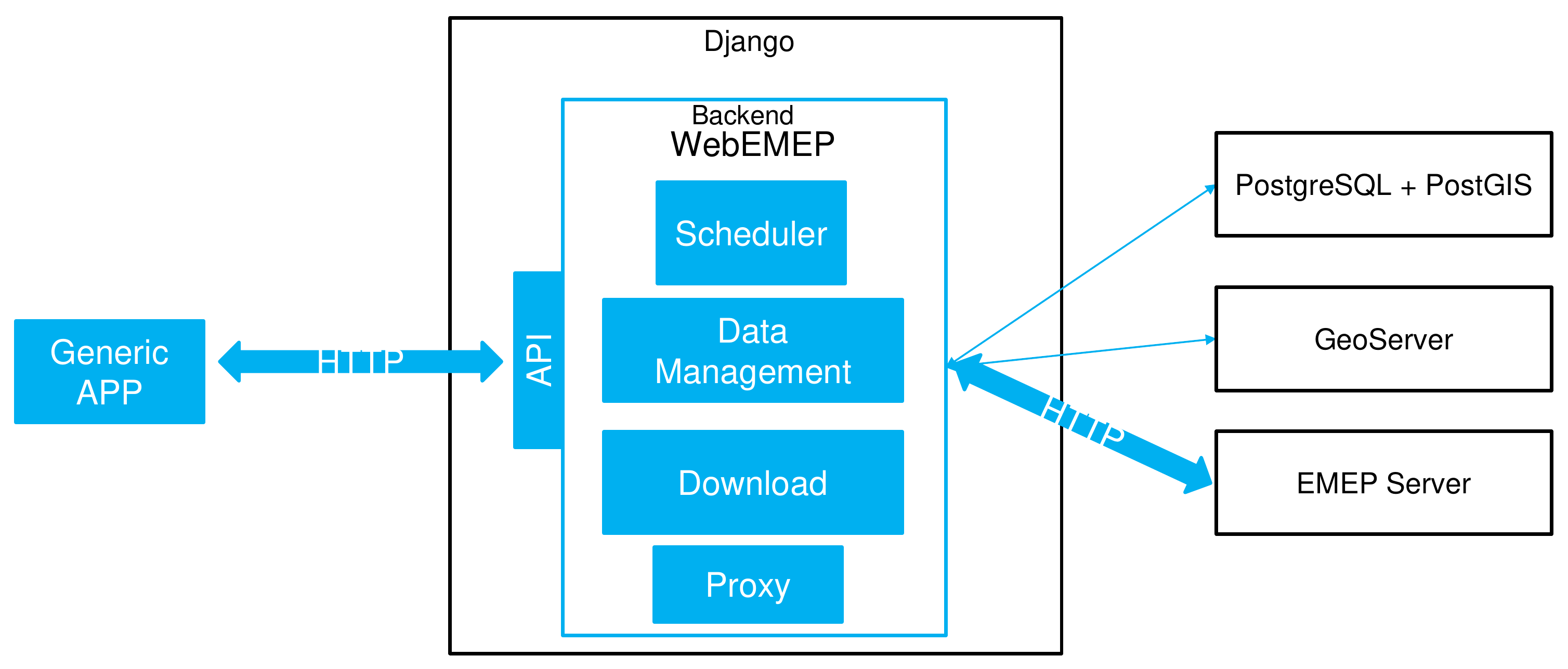}
    \caption{WebEMEP component in a generic Django framework, where blue boxes corresponds to WebEMEP, and the black bounded boxes have been developed by other entities.}
    \label{fig:WebEMEP_diagram}
\end{figure}

WebEMEP architecture is illustrated in figure \ref{fig:WebEMEP_diagram}. The backend part is responsible for extracting and management of the data. There is no user interface because all the mechanisms are automatic, including the mechanism to download the data.

The WebEMEP component is responsible for retrieving data from the EMEP site and to make the information available on the platform. The Norwegian Meteorological Institute provides a catalogue that contains all the available services and data sets, and their URL paths. Therefore, the first step is to download the catalogue. The next step is to download and store the data, using the catalogue. The data provided by EMEP represents all of Europe, but there is only interest in a portion of that data. So the downloaded data has to be cut by using a polygon feature delimiting the relevant spatial domain, which can be altered. After the cut, the data is uploaded to the GeoServer, using the GeoServer API.

The Norwegian Meteorological Institute updates their catalogue yearly, normally at the same time of the year. Thus, the best way to guarantee the use of model predictions made with the most recent EMEP model version is to use a scheduler which updates all available data on a day set by the programmer. The scheduler is created using the Celery, a Python Library that manages task queues. This scheduler is a periodic Task manager, that kicks off tasks at regular intervals. The tasks are then executed by available worker nodes in the cluster. The programmer can choose the date and the frequency of the task, that is responsible to replace all the data that is available on the EMEP server.

The data management uses a dictionary that contains information of the catalogue that helps to track the data stored on the GeoServer. In addition, the management of this data is made through the available OGC standards and/or through the user interface made available by GeoServer.

For security reasons, the GeoServer cannot be accessed directly from the outside. Therefore, a proxy that gives indirect access to the GeoServer is used,
that changes the URL and adjusts the query string of the request to match the OGC standards of the GeoServer. This proxy is responsible to forward the requests to the GeoServer or to the EMEP website, when the GeoServer is down. 


The GeoServer cannot be accessed directly from the outside. Therefore, it used a proxy that gives indirect access to the GeoServer. 
The API has only one available interface that uses the proxy to redirect to the GeoServer, changing the URL and adjusting the query string of the request to match the OGC standards of the GeoServer. This interface is used to get data from the GeoServer.

The WebEMEP component developed during this project is available in a git repository, \href{https://github.com/FMMFHD/WebEMEP}{https://github.com/FMMFHD/WebEMEP}.

\subsection{GeoNames}
\label{section:library:GeoNames}

\begin{figure}[!ht]
    \centering
    \includegraphics[width=7.5cm]{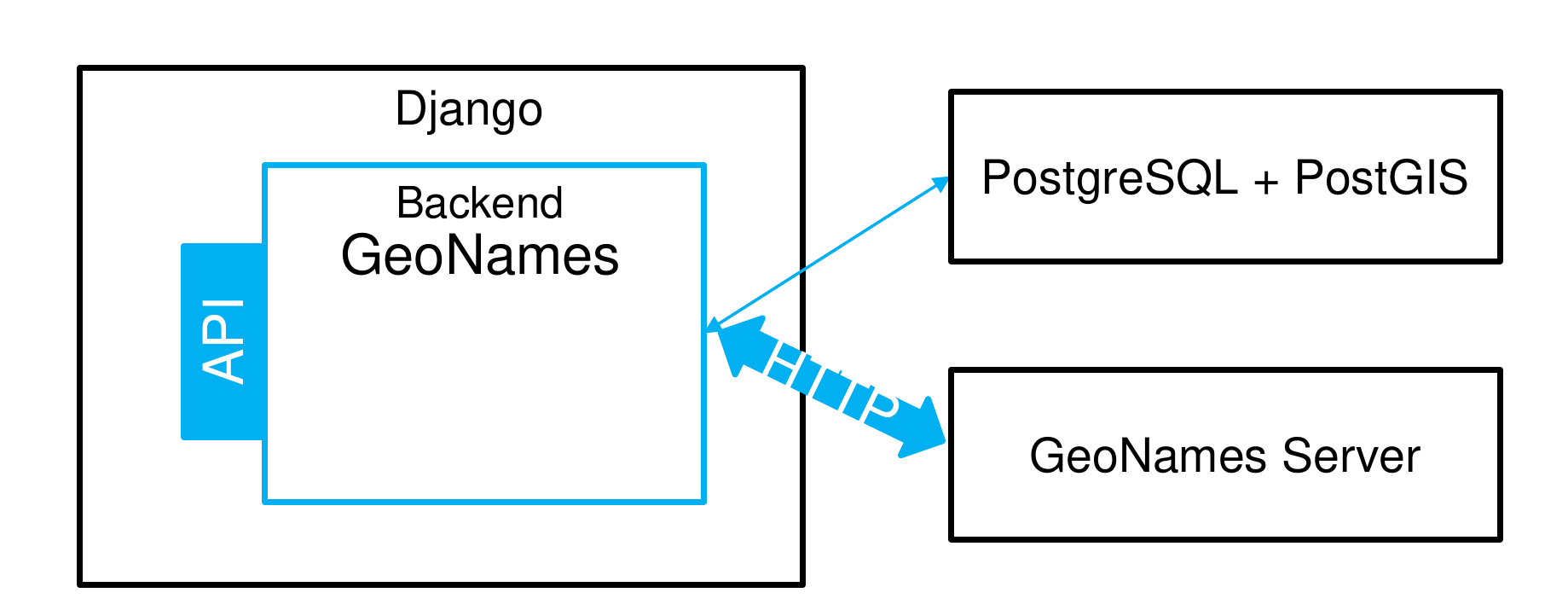}
    \caption{GeoNames component in a generic Django framework. Blue boxes corresponds to GeoNames, and the black bounded boxes have been developed by other entities.}
    \label{fig:GeoNames_diagram}
\end{figure}

The objectives of the GeoNames component is to facilitate the conversion between toponyms and coordinates. Besides storage, it offers access mechanisms for a framework where web applications with geospatial data are developed. GeoNames architecture is illustrated in figure \ref{fig:GeoNames_diagram}. The backend part is responsible for data download and for managing data access.

GeoNames is a geographical database that covers all countries and contains over eleven million toponyms. For each toponym it gives information about: latitude, longitude, population, etc.

The API has two access points. The first access point offers the possibility to auto complete sub-strings of toponyms. In other words, this feature will suggest toponyms that start with the same sub-string. The answer of the request is a list with the possible toponyms. The second access point offers the possibility to return all possible locations (several places may have the same name) for a given toponym. The answer of the request is a list with the all the possible coordinates.
    
The GeoNames component developed during this project is available in a git repository, \href{https://github.com/FMMFHD/GeoNames}{https://github.com/FMMFHD/GeoNames}.

\section{Demonstration Applications}
\label{sec:resul}

\begin{figure}[!ht]
    \centering
    \includegraphics[width=7cm]{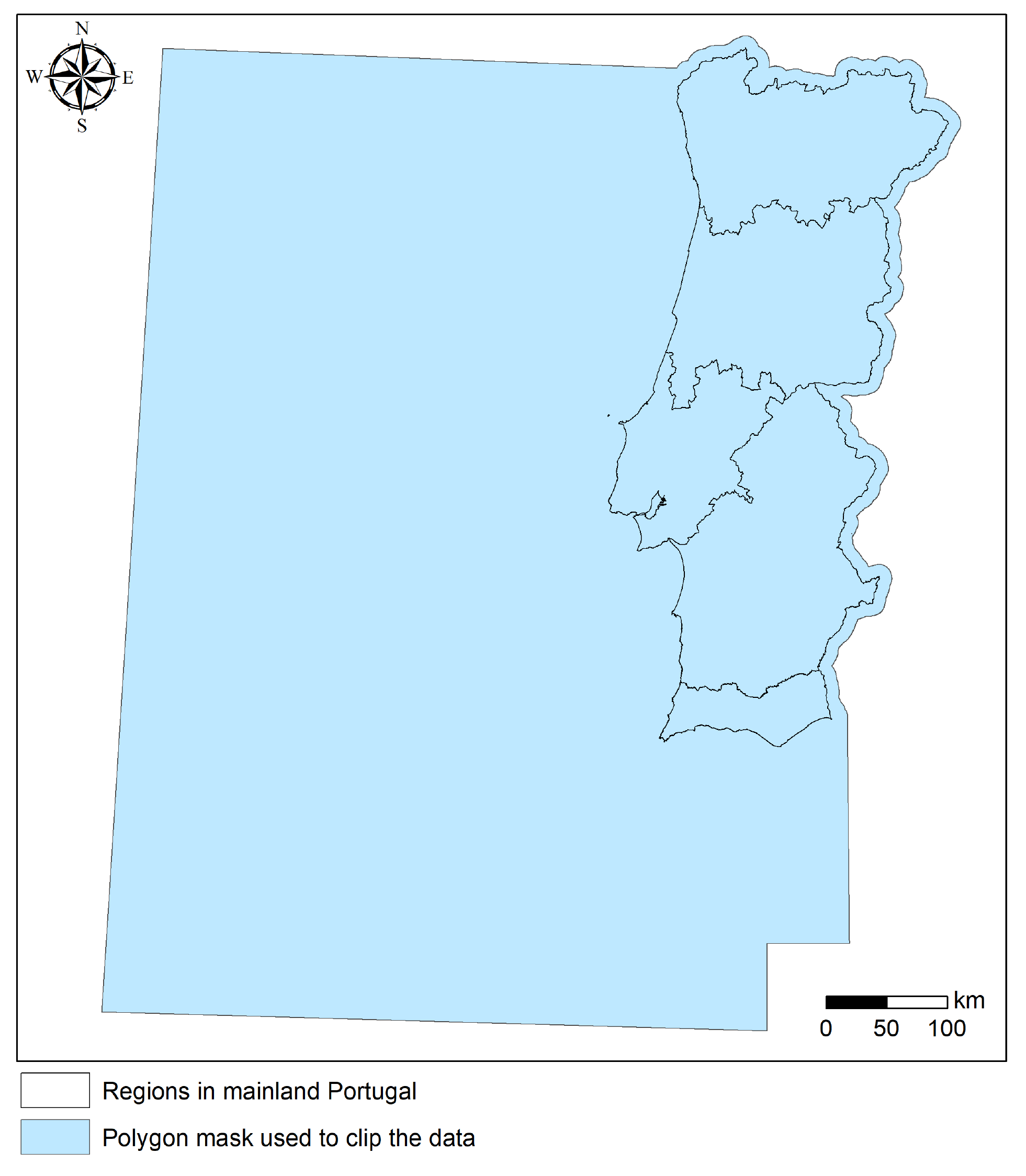}
    \caption{Mask used to clip and mask spatial data to use less memory. The 5 regions in mainland Portugal are also represented.}
    \label{fig:mask}
\end{figure}

The applications were developed with the specific aim to demonstrate results in mainland Portugal. For this reason, all data was clipped and masked with a vector file, uploaded to GeoServer, containing the region of interest, which includes a 10 km buffer around mainland Portugal and also the Portuguese margin up to a distance of approximately 400 km to the shore (figure \ref{fig:mask}).

The results presented in the applications require some data processing, such as determining temporal averages of observations, according to the desired temporal resolution, or regionally weighted averages, for EMEP model predictions for the main Portuguese regions (figure \ref{fig:mask}). Data processing required the extension of the WebAPA and WebEMEP components to include these methods. These methods comprise basic statistical data processing to aggregate temporal and spatial data, frequently used in geo-temporal data analysis.

\subsection{Evaluation of the EMEP model}
\label{section:applications:Validation}

One of the defined requirements was to evaluate the EMEP model using observations from APA's air quality stations. In this work, evaluation can be seen for mainland Portugal or for each of five regions, based on model predictions and observations falling within the polygon of interest.

So, the first step is to select the background air quality stations that fall within the selected polygon. 
Considering that from the restricted set of stations, for a given date and temporal resolution, only stations that captured more than 75\% of samples of the predicted number of captured samples (one sample per hour) are used. Due to these restrictions in spatial and attribute-based queries, there will be a limited number of stations to use. 
Each station is, then, used to extract EMEP model predictions. For every station the following evaluation is done.

The evaluation of the EMEP model is based on the following statistical measures recommend by Chang \& Hanna \cite{chang2004air, chang2005technical}: fraction of predictions within a factor of two of observations (FAC2), fractional bias (FB) and the normalised mean square error(NMSE).




According to the authors \cite{chang2005technical}, a model performs adequately or has "acceptable" performance when FAC2 $\geq$ 50\%, FB $\leq$ 30\% and NMSE $\leq$ 1.5. However, a model should not be excluded if one of the statistical measures fails to reach the indicative thresholds. For this reason Hanna and Chang \cite{hanna2011setting} suggest a comprehensive acceptance criterion of 50\%, meaning at least 50\% of these performance criteria are met, which means that two out of three statistical measures should be within the indicative thresholds.

\subsection{Dashboards}
\label{section:dashboard}

\begin{figure}[!ht]
    \centering
    \includegraphics[width=7.5cm]{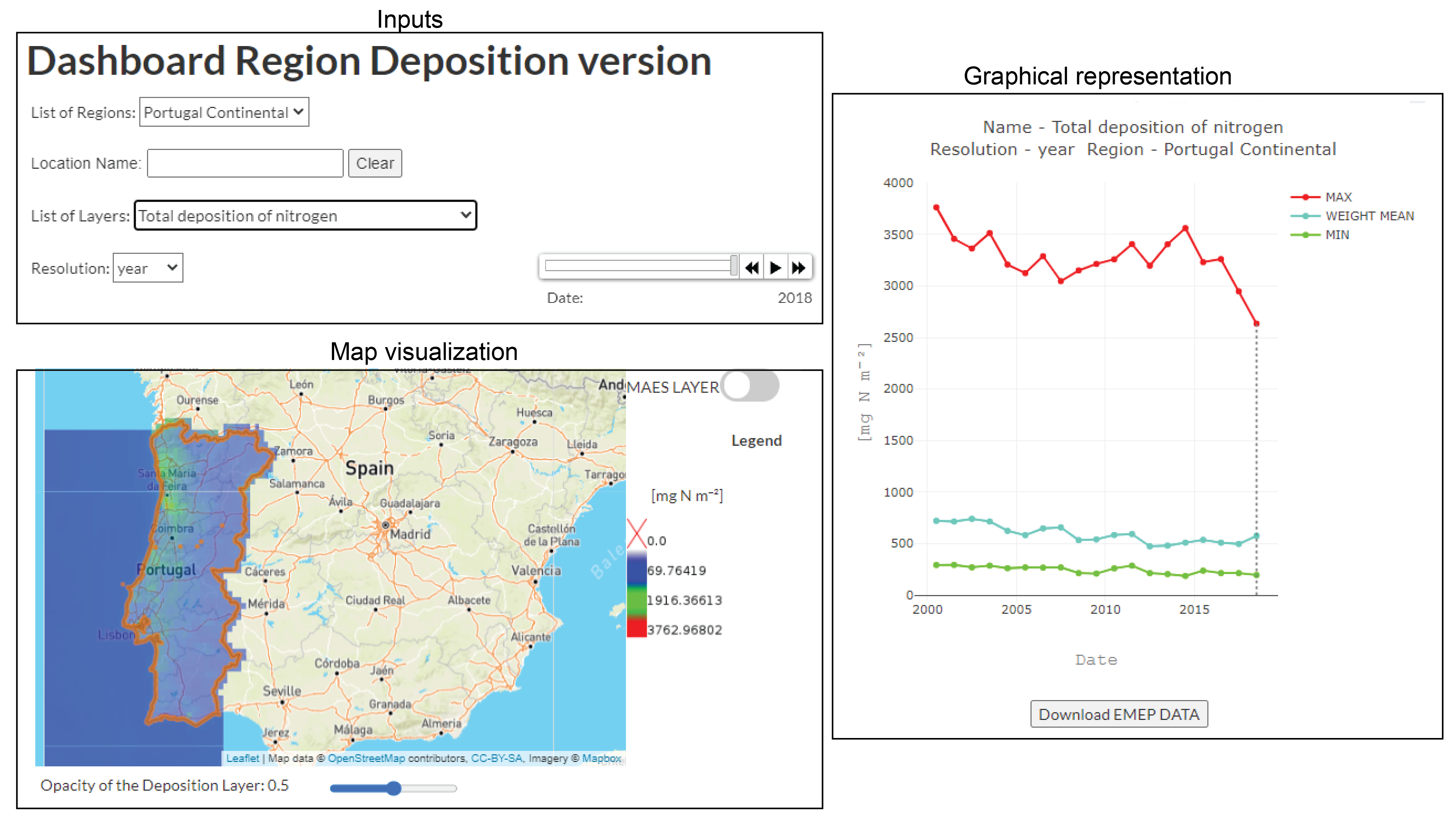}
    \caption{Deposition Dashboard Layout - inputs, map visualisation, graphical representation sections.}
    \label{fig:dashboard3}
\end{figure}

The Dashboards are an interactive web pages that gather and show official air pollution information for mainland Portugal. The web pages are divided in three sections: inputs, map visualisation and graphical representation (figure \ref{fig:dashboard3}).

\begin{figure}[!ht]
    \centering
    \includegraphics[width=7.5cm]{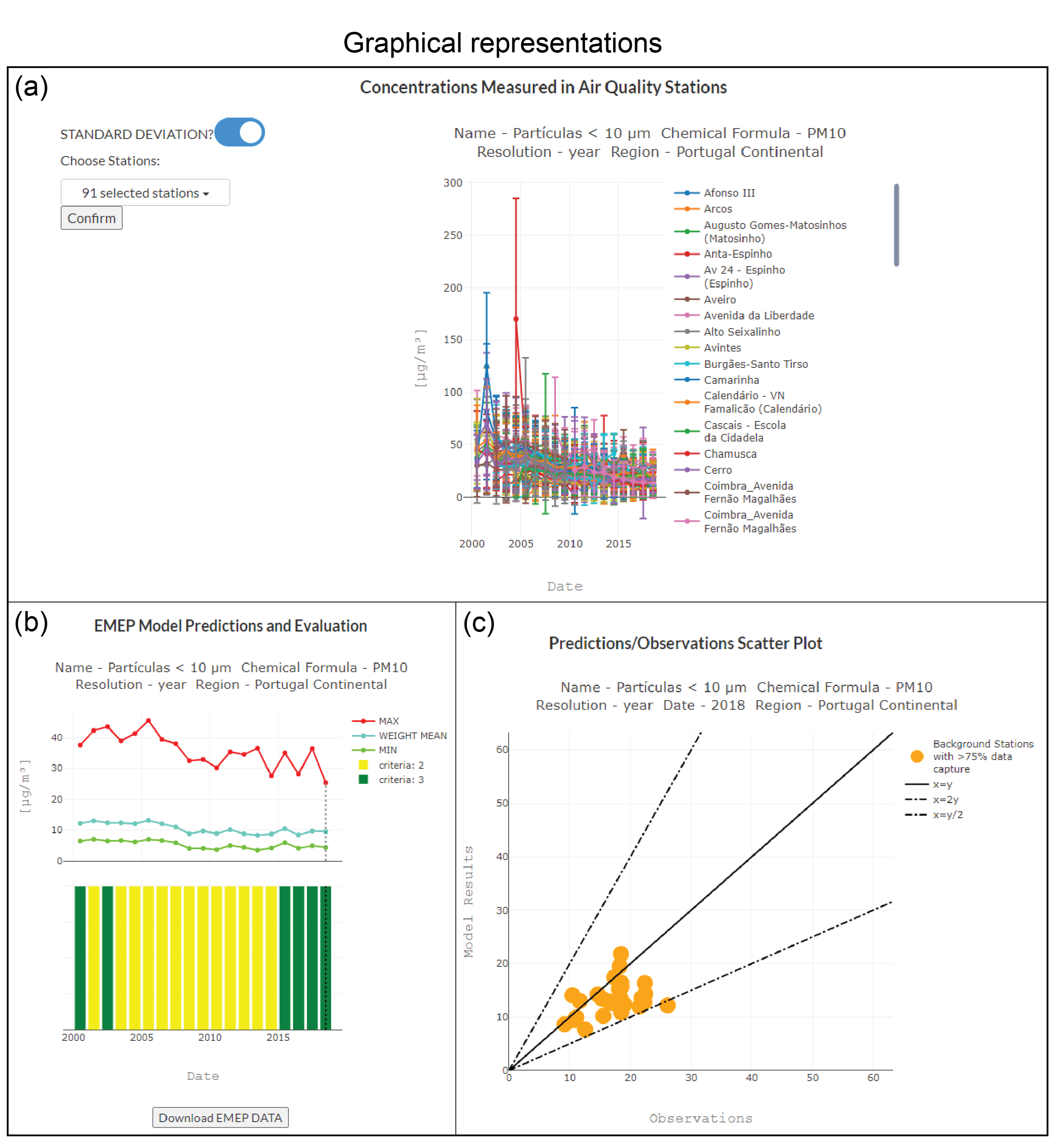}
    \caption{Concentration Dashboard Layout - graphical representation section.}
    \label{fig:dashboard2}
\end{figure}

Two Dashboards were developed, differing in the data presented, one showing data concerning modelled and measured concentrations of pollutants in the air, the other concerning only modelled nitrogen and sulphur deposition. The Dashboards developed are: 
\begin{itemize}
    \item Concentration Dashboard (figures \ref{fig:dashboard_layout},  \ref{fig:dashboard2}) - Data sets include both observations (measured in air quality stations), and EMEP model prediction of pollutant concentrations. In addition, land occupation/use classification is provided as an optional cartographic base. Furthermore, and exclusively for ammonia (NH\textsubscript{3}), nitrogen oxides (NO\textsubscript{X}), and sulphur dioxide (SO\textsubscript{2}), an optional representation of the spatial distribution of critical level exceedances is also provided.
    
    \begin{figure}[!ht]
        \centering
        \includegraphics[width=7.5cm]{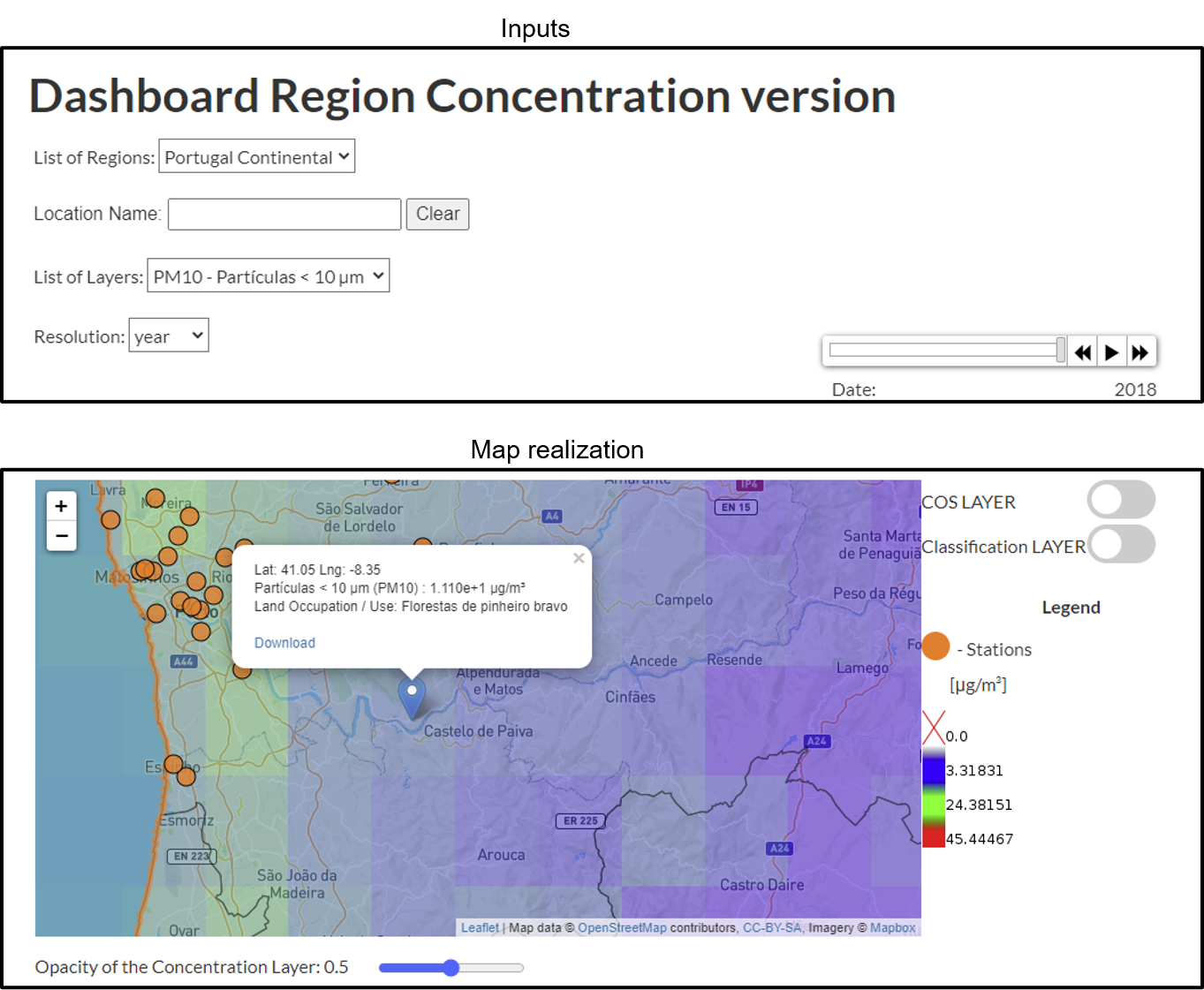}
        \caption{Concentration Dashboard Layout - inputs and map visualisation sections.}
        \label{fig:dashboard_layout}
    \end{figure}
    
    \item Deposition Dashboard (figure \ref{fig:dashboard3}) - Data set only includes EMEP model predictions of wet, dry and total deposition of nitrogen (oxidised and reduced forms) and sulphur. In addition, ecosystem classification is provided as an optional cartographic base. Furthermore, nitrogen critical load exceedances are also provided as an optional spatial representation, indicating whether the ecosystems are likely being affected, or not, by the amount of deposited nitrogen.
\end{itemize}

In the input section, there are five inputs: List of Regions, that comprises a select list allowing to choose between mainland Portugal or one of five regions in Portugal (\textit{Norte}, \textit{Centro}, \textit{Lisboa e Vale do Tejo}, \textit{Alentejo}, and \textit{Algarve}); Location Name, that is an input text that allows to type a location, or toponym; List of Layers, that comprises a select list, where the user can choose the layer of interest; Resolution, that comprises a select list where the user can choose the temporal resolution; and Date Selection, that comprises a slider, where the user can choose the date of interest. 
 
The next section in the Dashboards is the map visualisation. The Mapbox API \cite{MapBox_API} is the default cartographic base, using OpenStreetMap \cite{OpenStreetMap} as data source, and showing physical elements in the landscape, such as terrain (using hill-shade) and water bodies.
The detail of the cartographic information shown increases with map zoom. 


EMEP model predictions are presented as a layer, 
allowing the representation of the model predictions across the region of interest in an intuitive way, using a colour scheme, with red representing higher values and blue lower values, as shown in the legend (figure \ref{fig:dashboard_layout}).

By clicking on the map, or, alternatively, by typing a location in the input section, a blue marker appears, associated with a white popup (figure \ref{fig:dashboard_layout}) showing the following information: the coordinates of the marker, Latitude and Longitude; the name of the pollutant, its chemical formula, and the value of predicted concentration/deposition on that specific location; relevant cartographic information related with the content of the Dashboard (land-use classification in the Concentration Dashboard and ecosystem type classification on the Deposition Dashboard); location name, that only appears when typed in the input section; and the link to download the temporal variation of the pollutant in that location, extracted from the EMEP model prediction data set, that appears in the bottom of the popup.

In the map visualisation section, there is a slider, that allows to change the opacity of the pollutant layer, or the level of transparency that varies between 0 and 1.

In addition to the map visualisation section, the Dashboards offer data processing results in graphical format, that vary with selected inputs. The graphical representation section (figures \ref{fig:dashboard3}, \ref{fig:dashboard2}) is divided in 3 parts:

\begin{itemize}
    \item Concentrations Measured in Air Quality Stations Graph (figure \ref{fig:dashboard2}a) - A graph with time variation of the observation data. Only available in the Concentration Dashboard;
    
    \item EMEP Model Predictions Graph (figure \ref{fig:dashboard2}b) - A graph with temporal variation of spatially aggregated statistics (minimum, maximum and weighted mean) for the region of interest selected in the input section. In the Concentration Dashboard, a model evaluation chart also appears. Another particularity of this graph is that it allows to download the information in table form;
    
    \item Predictions / Observations Scatter Plot (figure \ref{fig:dashboard2}c) - A graph that compares observations with model predictions. Also exclusively available in the Concentration Dashboard;
\end{itemize}

\section{Evaluation}
\label{chapter:evaluation}

\subsection{Requirements Evaluation}
\label{chapter:evaluation:Rq_Verification}
The requirements described in section \ref{section:library:requirements} were met and demonstrated through the development of dashboards and the availability of the libraries developed.

\subsection{Usability evaluation}
\label{chapter:evaluation:Usability_evaluation}

Usability refers to the quality of a user’s experiences when interacting with a product. This evaluation was done through the survey using the method SUS (System Usability Scale) \cite{system-usability-scale}. This survey was answered by 10 researchers, getting a score of 85 out of 100.

\subsection{Utility evaluation}
\label{chapter:evaluation:Utility_evaluation}

The utility survey serves to assess the usefulness of the developed dashboards.  This survey was presented to users that will be more likely using the dashboards in the future: environmental/ecology researchers. 

The answers show that only one respondent out of 10 thinks that dashboards do not facilitate his/her work. 50\% of respondents think the data is presented in a simple way. On the other hand, only 50\% of respondents had previously accessed the data presented on the Dashboards by other means.

\subsection{Performance Evaluation}
\label{chapter:evaluation:Performance}
Although it has not been possible to measure the presentation times of the graphical interfaces of the developed applications, it can be considered that the presentation times are not very high, due to the usability results.

The second possible evaluation is related to the time the system takes to process the available data. The table \ref{table:1} indicates execution times to calculate the temporal variation of the EMEP Model Predictions, the temporal variation of the observations, and to evaluate the model. The time period under analysis was between 2000 and 2018. The applications were run in the background, on a server with 16GB of RAM and 64-bit processor. The applications run in the background to ensure that the results are available as quickly as possible.

\begin{table}[htb]
    \centering
    \begin{tabular}{|c|c|c|} 
         \hline
          & \multicolumn{2}{|c|}{resolution} \\
         Type of Results & Annual & Monthly \\ [0.5ex] 
         \hline\hline
         EMEP Model Predictions & $\approx$ 37,5 h & $\approx$ 19 days \\ 
         Observations & $\approx$ 6h & $\approx$ 2 days \\
         EMEP Model Evaluation & $\approx$ 20h & $\approx$ 9 days \\
         \hline
    \end{tabular}
    \caption{Execution times to calculate the temporal variation of the EMEP Model Predictions, the temporal variation of the observations, and to evaluate the model. The time lag was 19 years.}
    \label{table:1}
\end{table}

The execution times for the observations were the shortest, because the calculation used the average, maximum and minimum operations offered by the database, and these operations are efficient when applied to a single table.

To evaluate the model, it was necessary to calculate the data capture of all stations, this being the slowest operation in this process. The overlapping is a relatively fast operation because it uses the properties of the raster structure (matrix) to discover the pixel index to which the station to compare, belongs to. 

The process to determine weighted averages of EMEP model predictions is the slowest, because it needs to determine which pixels belong to the region of interest. To determine which pixels to consider is a slow process, being always necessary to verify if the region's polygon intersects the vectored pixel polygon. 

In addition, the time differences between annual and monthly resolution directly depends on the amount of data to be processed. The data with monthly resolution is 12 times larger than with annual resolution, and execution times differ approximately 12 times. For example, the execution times for processing EMEP model predictions are 37.5h for annual resolution, and 19 days (456h) for monthly resolution, the ratio being approximately 12.

By adding a new year of data, the system needs to process this data in both time resolutions (annual and monthly), taking approximately: 2h for Observations; 1 day for EMEP Model Predictions; and 12h for EMEP Model Evaluation. The new data is processed in the background, not affecting the functioning of the system, i.e. the system continues to provide results for the other years.

\section{Conclusions} \label{conclusion_section}
\subsection{Achievements}
\label{section:achievements}

The architecture of the libraries worked according of the requirements of gathering, processing and representing pollution data. 

The Dashboards were created because there is a need to automate processes related to air pollution data. To help the development of the Dashboards, four libraries have been developed, where two of them are responsible for collecting data from the APA (observations) and EMEP (model results). The other two libraries were created because of the need to facilitate the loading of excel files into the database and to facilitate translation between the toponym and its coordinates using the Geonames database.

There has been a positive assessment of the dashboards by researchers who have evaluated it. Through the surveys, the dashboards were evaluated as 85 out of 100 in the system usability scale (SUS), showing that they are relatively easy to use and show results that can be useful in air pollution studies.

For data with annual resolution, it took about 1.5 days to process all data automatically, which, through the cache, becomes available to users in a matter of seconds. If researchers had to process this data using commonly used commercial and open source applications (e.g., R, GIS software), it would take much longer and human intervention (prone to errors) would be required between each step. For monthly resolution, where the data is much larger, the whole processing took about two weeks.
It should be noted that the times mentioned before corresponds to a process of 20 years of data, in the case of one year, it takes no more than one day to process all resolutions.

With these applications, it has been possible to solve collection and processing problems for researchers. In the future, depending on their needs, it is possible to develop new libraries for new types of data (e.g., air pollution derived from satellite data), using the same architecture, and to create new applications to present new results in a simple way.


\subsection{Future Work}
\label{section:Future_Implementations}

In the future, it would be interesting to implement suggestions given by specialists through the survey.

By clicking on the dashboard map, there is information about the ecosystem or about land-use. But the downloaded data of that point does not include this information at the moment. To implement this feature, it would be only necessary to know the coordinates that are given by the request parameters and ask through the WFS standards to the geospatial data server. After getting the response, it would be necessary to check which polygon the point belongs to. Having the polygon, the final step would be to add the information to the download data. Another improvement of the graphical interface would be the possibility of drawing a polygon on the map to replace the regions available. Due to the available plugins in the Leaflet library, it is possible to draw polygons on the map. Having a drawn polygon, it is simple to extract its coordinates and send them to the server. Once having the coordinates limiting the polygon, the server can then start to process the data, following the same procedure as when calculating for a predefined polygon from a region in mainland Portugal. Therefore, it is possible to implement this functionality. The drawback is the time that the user has to wait for all the process to be complete, because the server does not have the capability to store all results, for all possible polygons, on a given map. The time consumption increases with the size of the drawn polygon.

One of the improvements to be made in the developed libraries is to have available on the Dashboards multiple versions of the downloaded EMEP prediction model. Therefore, there would be no need to replace the data when there is a new update. There is no evaluation of model data on the Deposition Dashboard. Therefore, it is necessary to evaluate this data, to give some confidence to the results presented in the dashboard.

It would be interesting to integrate new data such as: natural habitat information; climate information; and satellite information. Natural habitat information would be useful to researchers because it identifies the area where a particular species lives. This information has the same type as ecosystem information, so the only thing to do is to download and store this information. After the data is stored, the use of this data is the same as for MAES (ecosystem) data, already developed. The pollution is related with climate, because the weather influences how pollution is transported in atmosphere and/or it is deposited. So, a future implementation would be to provide this information to the user. Climate information (spatial and temporal data of precipitation, temperature, wind) is available as raster just like EMEP data, so the data procedure is similar. The difficult is on the representation of the wind on the map, because the wind direction is not instinctive to read with colour graduation, but should be represented with arrows (e.g. one arrow indicating the wind direction per pixel). The precipitation and temperature data are scalar, and should be represented on the map with a colour scale just like the EMEP model predictions. Satellite data can be used to evaluate EMEP model results. The procedure is the same using satellite data, but instead of calculating the valid stations, the valid regions would be calculated.

Although there was a large number of suggestion that are relevant  and that could provide relevant results, we think that only a set of dashboards and backend infrastructures like ours could allow the successful implementation of such suggestions, thus demonstrating the need and innovation of the work described in this thesis.

\bibliographystyle{plain}
\bibliography{references}

\end{document}